\documentclass[twocolumn, superscriptaddress,showpacs,preprintnumbers,amsmath,prl]{revtex4}

\usepackage{graphicx}
\usepackage{dcolumn}
\usepackage{bm}

\usepackage[normalem]{ulem}

\usepackage{color}
\definecolor{green}{rgb}{0.0, 0.5, 0.0}

\usepackage{ulem}

\begin{document}
\preprint{APS/123-QED}

\title{Tuning the magnetic dimensionality by charge ordering in the molecular TMTTF salts}

\author{Kazuyoshi Yoshimi}
\affiliation{Department of Physics, University of Tokyo, Tokyo 113-8656, Japan}
\affiliation{Nanosystem Research Institute ``RICS," AIST, Ibaraki 305-8568, Japan}
\author{Hitoshi Seo}
\affiliation{Condensed Matter Theory Laboratory, RIKEN, Saitama 351-0198, Japan}
\affiliation{JST, CREST, Saitama 351-0198, Japan}
\author{Shoji Ishibashi}
\affiliation{Nanosystem Research Institute ``RICS", AIST, Ibaraki 305-8568, Japan}
\author{Stuart E. Brown}
\affiliation{Department of Physics and Astronomy, UCLA, Los Angeles, California 90095, USA}
\date{\today}

\begin{abstract}
We theoretically investigate the interplay between charge ordering and magnetic states in quasi-one-dimensional molecular conductors TMTTF$_2X$, motivated by the observation of a complex variation of competing and/or coexisting phases. We show that the ferroelectric-type charge order increases two-dimensional antiferromagnetic spin correlation, whereas in the one-dimensional regime two different spin-Peierls states are stabilized. By using first-principles band calculations for the estimation for the transfer integrals and comparing our results with the experiments, we identify the controlling parameters in the experimental phase diagram to be not only the interchain transfer integrals but also the amplitude of the charge order. \end{abstract}
\maketitle

Low-dimensional molecular conductors provide a fruitful stage to study strong electron correlations
in the presence of large quantum fluctuations and coupling to lattice degrees of freedom~\cite{review}. The observed phase transitions involving spin, charge, and lattice degrees of freedom are summarized in the form of pressure ($P$)-temperature ($T$) phase diagrams for different families. It is common to stabilize a different ground state even for relatively small pressure variations. Presumably, these changes are triggered by small variations in lattice constants for a given material while maintaining the same geometry of constituent molecules at room $T$ (isostructural)~\cite{Seo04}.A general goal is to identify the parameters controlling the ground states and trends in the nature of the elementary excitations. An example of interest is the TMTTF$_2X$ (TMTTF = tetramethyltetrathiafulvalene; $X$, monovalent anion) family of molecular solids~\cite{review2}, in which the tuning of charge order, by way of applying pressure, appears to play a role in controlling the magnetic states~\cite{Yu04}. Specifically, decreasing the charge order amplitude by the use of applied pressure is associated with the antiferromagnetic (AFM) transition $T_{\rm N}\to 0$, and clarifying the relevant physics for this behavior is of interest in the field of quantum magnetism.

\begin{figure}[bt]
\begin{center}
\resizebox{75mm}{!}{\includegraphics{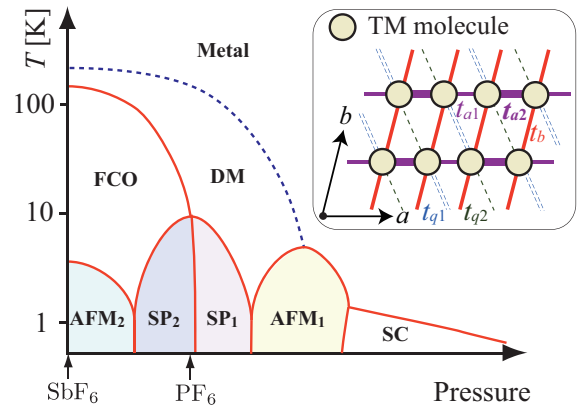}}
\end{center}
\vspace*{-2em}
\caption{(Color Online)
A schematic phase diagram for TM salts~\cite{Yu04}.
The ambient pressures for TMTTF$_2$SbF$_6$ and TMTTF$_2$PF$_6$ are shown.
DM, FCO, AFM, SP, and SC represent dimer-Mott, ferroelectric charge ordering,
 antiferromagnetic, spin-Peierls, and superconducting states, respectively.
The dashed line is a crossover while the solid lines are phase transitions.
The inset shows the arrangement of TM molecules in the conducting plane. 
}
\label{fig1}
\end{figure}
In the quasi-one-dimensional family of 
TM$_2X$ (TM: TMTTF or tetramethyl-tetraselenafulvalene=TMTSF),  
the key parameter has been widely accepted as the dimensionality ($D$) tuned by the relative increase of interchain transfer integrals by $P$~\cite{Jerome91,Emery82, review2}. Figure~\ref{fig1} shows the recently updated phase diagram~\cite{review2, Chow00, Yu04, Zamborszky02}. Amazingly, a wide variety of phase transitions appear by applying $P$ or a replacement of $X$ (chemical $P$). The phases latest revealed are in the left side where a ferroelectric-type charge ordering (FCO) transition was found~\cite{Chow00,Monceau01}; it has a strongly correlated nature~\cite{Seo06}, leading to magnetic transitions at low $T$. Prior to the discovery of the FCO phases, the $P$-$D$ correspondence was invoked to describe the phases and transitions appearing at higher $P$: Transport experiments indicate that the correlation gap 
is reduced with $P$, interpreted as driven by the transverse hopping process~\cite{Vescoli98}, 
and the system shows a dimensional crossover~\cite{review2}. 
The low-$T$ spin-Peierls (SP) state (SP$_1$ in Fig.~\ref{fig1}) is destabilized in favor of an AFM state (AFM$_1$)~\cite{Chow98}, 
which is consistent with the increase in 
the transverse spin-exchange couplings~\cite{Inagaki83,Schmeltzer99}.

Difficulties come about when one attempts to apply the relation to the left side, the region with FCO.
First, a discrepancy is easily seen since another AFM phase (AFM$_2$) appears at the lowest $P$
and turns into the SP phase (SP$_2$) by applying $P$, opposite to the SP$_1 \rightarrow$ AFM$_1$ variation; the AFM$_2$ and SP$_2$ states both coexist with FCO. Another point is that the FCO and AFM$_2$ transitions behave cooperatively; namely, their transition temperatures both develop at low $P$~\cite{Yu04, Iwase10}. This is peculiar in the sense that, in general, charge ordering tends to decrease the effective spin-exchange couplings~\cite{Seo06, Tanaka05} and, therefore, would diminish transitions subject to spin couplings; in fact, the SP$_2$ phase shows such behavior~\cite{Chow00,Zamborszky02}, which is reproduced in theoretical works~\cite{Otsuka05, Clay03}.

In this Letter, we theoretically elucidate the origin for such variations which apparently do not fit to the conventional practice. Starting by evaluation of transfer integrals using first-principles band calculations (FPBC), we then investigate the roles of electronic correlation and electron-lattice couplings on the basis of the effective quarter-filled extended Hubbard model (EHM). We will show that the complex sequence of phases observed experimentally can be reproduced naturally when we add the intersite Coulomb interaction as another essential parameter, in addition to the interchain transfer integrals.

The inclusion of the newly found phases in the low-$P$ side of the phase diagram was proposed based on NMR measurements~\cite{Yu04, Zamborszky02}, and the continuous connection of phases has been shown by different experiments~\cite{Klemme95,Adachi00,Jaccard01,Itoi08}.First we address this in terms of electronic structure. We calculate the electronic band dispersions for two (TMTTF)$_2X$ members situated in the FCO region, $X$=PF$_6$ (SP$_2$ phase) and $X$=SbF$_6$ (AFM$_2$ phase), within FPBC by using the computational code QMAS (Quantum MAterials Simulator)~\cite{QMAS} based on the projector augmented-wave method~\cite{PAW} with the generalized gradient approximation~\cite{GGA}.

By tight binding fitting to the electron bands near the Fermi level we obtain the values of transfer integrals in the unit of meV as $\{t_{a1}, t_{a2}, t_{b}, t_{q1}, t_{q2} \}$$=$$\{-155, -203, 26.2, -1.31, -3.29\}$  for the former and $\{-149, -207, 16.4, -16.4, -9.73\}$ for the latter salt (notations are shown in the inset in Fig.~\ref{fig1}). The absolute values of $t_{a1}$ and $t_{a2}$ are about 10 times larger than the other transfer integrals: Both salts form a quasi-one-dimensional electronic structure along the $a$ axis with dimerization. A measure for the $D$ effect is given by $|t_b/t_{a2}|$ whose values are given as $0.129$ for the PF$_6$ salt and $0.080$ for the SbF$_6$ salt. From this point, as far as the transfer integrals are concerned, the 2-$D$ in the PF$_6$ salt is indeed higher than that in the SbF$_6$ salt. This is consistent with the semiempirical extended H\"uckel calculations as well as considerations based on their crystal structures~\cite{Xray_data}.

Next we investigate the role of Coulomb repulsions on top of such an electronic structure, by considering the quasi-one-dimensional EHM at quarter-filling in terms of holes. The Hamiltonian is given by
\begin{align}
&{\cal H}_{\rm EHM}=-\!\sum_{\langle i j\rangle, \sigma} t_{ij}(c_{i\sigma}^{\dag}c_{j\sigma}+{\rm H.c.})\nonumber\\
& \hspace{20mm} +U\!\sum_{i}n_{i\uparrow}n_{i\downarrow} +\sum_{\langle ij\rangle}V_{ij}n_{i}n_{j},
\label{EHM}
\end{align}
where $t_{ij}$ is the transfer integral between the neighboring sites denoted by $\langle i j \rangle$, $c_{i \sigma}^{\dag}$ ($c_{i \sigma}$) is the creation (annihilation) operator of a hole on the $i$th site with spin $\sigma=\uparrow$ or $\downarrow$, and  $n_{i}=n_{i\uparrow}+n_{i\downarrow}$ with $n_{i\sigma}=c_{i\sigma}^{\dag}c_{i\sigma}$. $U$ and $V_{ij}$ are the on-site and the intersite Coulomb interactions, respectively. From the results of FPBC, we hereafter set the transfer integrals as $t_{a1}=-0.8,~t_{a2}=-1,$ and $t_{q1}=t_{q2}=0$~\cite{Work_SF} and choose the interchain transfer integral $t_b$ as a parameter, as inferred from the results above. 
We choose the on-site Coulomb interaction to be a typical value for this class of materials~\cite{Seo06}, as $U=4$ ($\sim 1$ eV ) and impose a constraint on $V_{ij}$ as $V_{a1}=V_{a2}=V_{q1}=V_{q2}=V$ and $V_b=0$ to realize the FCO pattern observed in experiments (see Fig.~\ref{fig3}).

Numerical exact diagonalization on a $4\times4$ sites cluster under periodic boundary conditions is performed, where we introduce interdimer or intradimer charge and spin structure factors given by $C_{\pm}(\bm q)= N_d^{-1}\sum_{i, j}\langle n_{i}^{\pm}n_{j}^{\pm}\rangle {\rm e}^{i {\bm q}\cdot({\bm r}_i-{\bm r}_j)}$ and $S_{\pm}(\bm q)= N_d^{-1}\sum_{i, j}\langle m_{i}^{\pm}m_{j}^{\pm}\rangle {\rm e}^{i {\bm q}\cdot({\bm r}_i-{\bm r}_j)}$, respectively, where $N_d$ is the total number of dimers and ${\bm r}_i$ denotes the center position of the $i$th dimer. Here, the interdimer($+$) [intradimer($-$)] correlations are detected by the summation [difference] in charge and spin densities within each dimer,
$n_{i}^{\pm} =(n_{2i}\pm n_{2i+1})/2$ and $m_{i}^{\pm} =(m_{2i}\pm m_{2i+1})/2$ with $m_{i}=n_{i\uparrow}-n_{i\downarrow}$, respectively, where the even (odd) number is labeled as the site for the left (right) side in a dimer.

\begin{figure}[t]
\begin{center}
\resizebox{85mm}{!}{\includegraphics{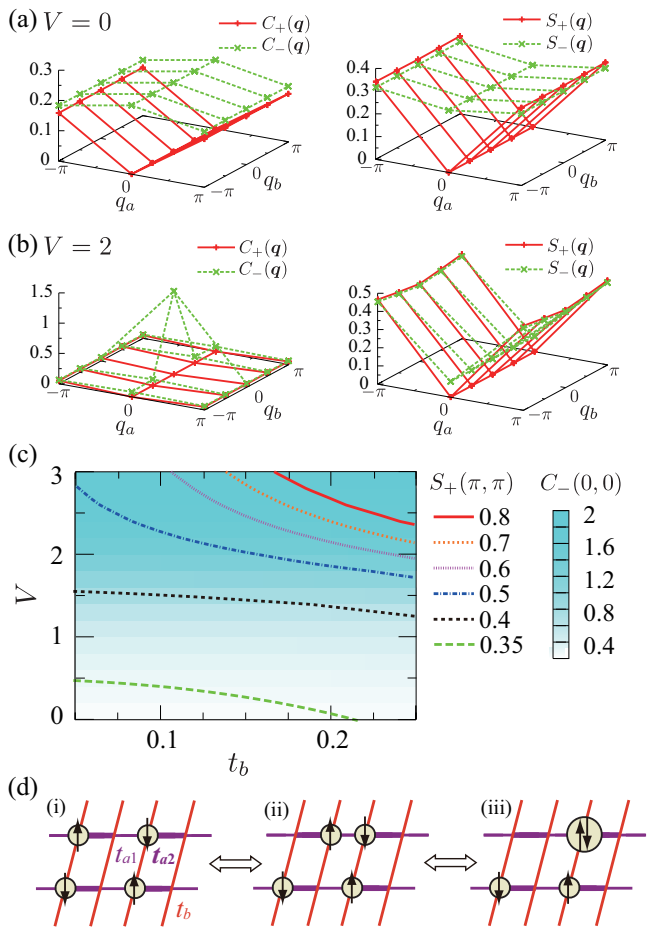}}
\end{center}
\vspace*{-2em}
\caption{(Color Online) Interdimer and intradimer charge and spin structure factors $C_\pm ({\bm q})$ and $S_\pm ({\bm q})$ at  (a) $V=0$ and (b) $V=2$, for $U=4$ and  $t_b=0.1$. (c) $C_{-}(0, 0)$ (background color) and $S_+ (\pi, \pi)$ (contour) on the $(t_b, V)$ plane: The two-dimensional AFM correlation is developed by both $t_b$ and $V$. (d) The leading spin-exchange process from the fourth-order perturbation in the presence of FCO.
}
\label{fig2}
\end{figure}

Figure~\ref{fig2} shows $C_{\pm} ({\bm q})$ and $S_{\pm} ({\bm q})$ for $t_b=0.1$, at $V=0$ (a) and $V=2$ (b)\cite{comment_for_V}. At $V=0$, there is no pronounced peak in $C_{\pm} ({\bm q})$; the system is in the dimer-Mott (DM) insulating state, where the intrachain dimerization together with $U$ leads to a Mott insulator~\cite{Seo06}. The enhanced $S_{+} (\pi, q_b)$ and featureless behavior in $S_{-} ({\bm q})$ indicate that the AFM correlation is developed between dimers, but only in the $a$ direction due to the 1-$D$. On the other hand, at $V=2$, $C_{-}({\bm q})$ has a clear peak at ${\bm q}=(0, 0)$, i.e., the FCO correlation, and $S_{\pm} ({\bm q})$ both have peaks at ${\bm q}=(\pi, \pi)$. This shows that the development of FCO due to the intersite Coulomb interaction induces the two-dimensional AFM correlation between the charge rich sites. Note that this happens in spite of the fact that the transfer integrals are unchanged from Fig.~\ref{fig2} (a).%

The emergence of 2-$D$ AFM correlation is balanced by the degree of FCO and the interchain transfer integral, as seen from Fig.~\ref{fig2} (c), where $C_{-}(0, 0)$ and $S_{+} (\pi, \pi)$ are plotted on the $(t_b, V)$ plane.
$C_-(0, 0)$ sharply develops with increasing $V$ while $S_+(\pi, \pi)$ increases with increasing $t_b$, expected from the interchain spin exchange.
The noticeable point is that the $S_+(\pi, \pi)$ peak is rapidly developed at large $V$ when FCO is stabilized: 
The FCO state assists in stabilizing the AFM state.
As a result, the one-dimensional regime
in the spin sector is limited to the region where {\it both} parameters $t_b$ and $V$ are small.%

The origin of the magnetic properties seen in Fig.~\ref{fig2}~(c) can be understood by a simplified strong-coupling analysis estimating the leading terms of the spin-exchange coupling by perturbation calculations with respect to the transfer integrals. In the DM state, the spin-exchange coupling between dimers along the $a$ axis is simply given by $J_a=-t_{a1}^2/U_d$, while that along the $b$ axis is given by $J_b=-4t_{b}^2/U_d$, where $U_d$ is the effective on-site Coulomb interaction for the dimer units~\cite{Seo06}. Thus, the 2-$D$ is enlarged toward $|t_{a1}|\sim 2t_b$, namely, $t_b\sim 0.4$; in fact $S_+(\pi, \pi)$ shows a maximum around this value for $V=0$ (the large $t_b$ region is not shown). 
On the other hand, in the basis of the FCO state in the limit of large $U$ and $V$, 
the charge localizes on every other site along the intrachain $a$ axis and on the nearest-neighbor sites along the $b$ axis (see Fig.~\ref{fig2}~(d)); 
then the spin-exchange coupling between these sites for the former is given by 
$J_a \sim -4 t_{a1}^2 t_{a2}^2 / (9 UV^2)$ from the fourth-order perturbation whose spin-exchange process is shown in Fig.~\ref{fig2}~(d)~\cite{Ohta94}, while for the latter $J_b\sim -4t_b^2/U$ from the second-order perturbation.
Although $t_b$ is small compared to $t_{a1}$ and $t_{a2}$, $J_b$ can become the same order compared to $J_a$ due to the effect of $V$. Then the 2-$D$ in the magnetic state increases and the AFM state is induced.

\begin{figure}[t]
\begin{center}
\resizebox{70mm}{!}{\includegraphics{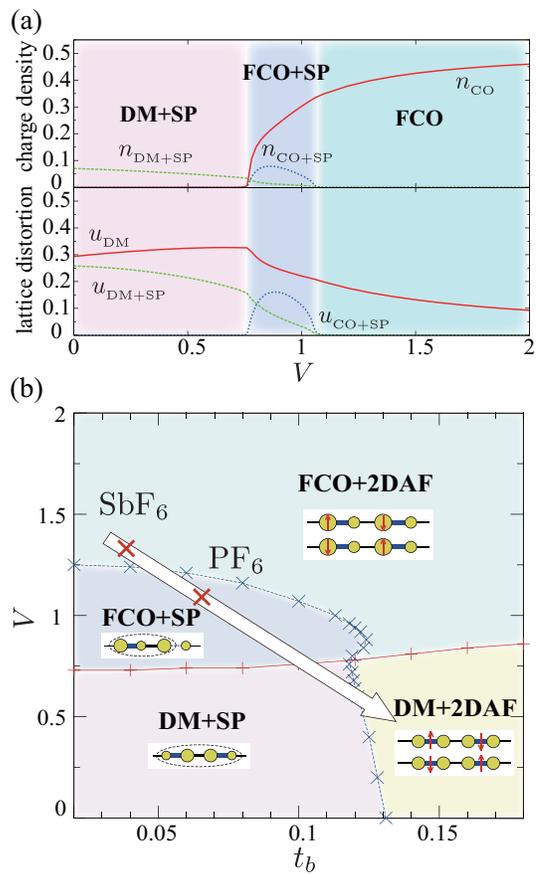}}
\end{center}
\vspace*{-2em}
\caption{(Color Online) (a) $V$ dependence of the order parameters in the charge densities and lattice distortions for $U=4$, $t_b=0.1$, $K_1=0.8$, and $K_2=1$. 
(b) Ground state phase diagram on $(t_b, V)$ plane for $U=4$, $K_1=0.8$, and $K_2=1$. 
``2DAFM'' stands for the enhanced two-dimensional AFM correlation 
while the other abbreviations (see text) represents ordered phases. 
The proposed trajectory for the pressure axis in the phase diagram Fig.~\ref{fig1} is shown by the arrow. 
}
\label{fig3}
\end{figure}%
It is known that electron-lattice couplings invoke the SP transition and various types of phase transitions with lattice modulations within one-dimensional models~\cite{Otsuka05, Clay03, Riera00}. Here we investigate such effects in our quasi-one-dimensional system by considering both Peierls- and Holstein-type electron-lattice interactions coupled to the EHM, given in addition to Eq.~(\ref{EHM}) as
\begin{align}
&{\cal H}_{\rm P} =-\!\sum_{\langle i j\rangle_a, \sigma} t_{ij} u_{ij}(c_{i\sigma}^{\dag}c_{j\sigma}+{\rm H.c.})+\frac{K_1}{2}\sum_{\langle ij\rangle_a}u_{ij}^2,\label{HP}\\
&{\cal H}_{\rm H} =-\sum_{i}v_i n_i+\frac{K_2}{2}\sum_{i}v_i^2,\label{Ham_Hol_phonon}
\end{align}
where $u_{ij}$ and $v_{i}$ are the renormalized lattice distortions treated here as classical values and the corresponding spring constants are given by $K_1$ and $K_2$, respectively.
The Peierls distortions are treated along the $a$ axis [written as $\langle i j\rangle_a$ pairs in Eq.~(\ref{HP})]: namely, only intrachain couplings are considered, to account for the one-dimensional quantum effects. Using the Hellman-Feynman theorem under the constraint $\sum_{\langle ij \rangle_a} u_{ij}=0$, we can obtain $u_{ij}$ and $v_i$ $(=\langle n_i \rangle/K_2)$ self-consistently by using the ground state expectation values for bond operators and charge densities~\cite{Riera00,Otsuka05, Clay03}.
In the following, we show results for an $8\times2$ site cluster~\cite{comment_size} at $K_1=0.8$ and $K_2=1$,
under the antiperiodic and periodic boundary conditions along the $a$ and $b$ axis, respectively.
From the four-lattice periodicity modulations along the $a$ axis, we can define order parameters by following Ref.~\cite{Otsuka05} as, for the FCO state, $n_{_{\rm CO}}$: for the coexistence of FCO and SP tetramerization (FCO+SP), $\{ u_{_{\rm CO+SP}}, n_{_{{\rm CO}+{\rm SP}}} \}$: and for the SP state without FCO (DM+SP), $\{u_{_{{\rm DM}+{\rm SP}}}, n_{_{{\rm DM}+{\rm SP}}}\}$~\cite{noteOP} [see Fig. \ref{fig3} (b) for schematic representations]. Because of intrinsic dimerization ($t_{a1},t_{a2}$), two fold lattice distortion $u_{_{ \rm DM}}$ always exists.

Figure \ref{fig3} (a) shows the results for $t_b=0.1$ (the same as Figs.~\ref{fig2} (a) and (b)),
as a function of $V$. As $V$ increases, first a phase transition occurs as DM+SP $\rightarrow$  FCO+SP
due to the effect of $V$~\cite{Otsuka05}
and then to the FCO state without SP tetramerization; the SP state becomes unstable by the development of the two-dimensional AFM correlation controlled by the FCO that we have seen above.
The ground state phase diagram on the $(t_b,~V)$ plane is shown in Fig.~\ref{fig3} (b).
The two kinds of SP states are suppressed with increasing $t_b$, due to the increase in the interchain spin exchange, while, as we have seen in Fig.~\ref{fig3} (a), $V$ also diminishes the SP states.
We confirm that both $S_{\pm}({\bm q})$ have sharp peaks at ${\bm q}=(\pm \pi, \pm \pi)$ in the FCO state, while only $S_+({\bm q})$ has peaks at ${\bm q}=(\pm \pi, \pm  \pi)$ in the DM state: These tendencies are the same for the case without electron-phonon couplings shown in Fig.~\ref{fig2}.

Based on the above results, we finally discuss our results in relation to
 the complex variation of phases in the phase diagram in Fig.~\ref{fig1}.
To establish a correspondence between $t_b$ in our calculation in Fig.~\ref{fig3}
 and our estimations based on FPBC mentioned above,
 we need to divide the latter by $2$ due to the $8\times 2$ cluster having a ``ladder" geometry;
 then for the PF$_6$ and SbF$_6$ salts, this gives $t_b=0.065$ and $t_b=0.040$. 
By considering the experimental ground states for the PF$_6$ salt (FCO+SP) and the SbF$_6$ salt (FCO+2DAFM), we can deduce that they are positioned as indicated in Fig.~\ref{fig3} (b), where the SbF$_6$ salt has larger value of $V$. This is consistent with the fact that the SbF$_6$ salt has larger transfer integrals along the diagonal $q1$ and $q2$ bonds, namely, larger overlap between the molecular orbitals, which results in larger values of intersite Coulomb repulsions~\cite{TMori00}, $V_{q1}$ and $V_{q2}$, favoring the FCO pattern.
The smooth evolution of phases with applied $P$ 
 suggests that the system follows along the arrow in Fig.~\ref{fig3} (b).
Specifically, with applied $P$, transfer integrals reflecting the overlap between the molecular orbitals 
are more sensitive compared to the inter-site Coulomb repulsions, which are approximately a function of intermolecular distance~\cite{TMori00}.
Then, the variation of ground state with $P$ is now given by
FCO+2DAF (AFM$_2$) $\to$ FCO+SP (SP$_2$) $\to$ DM+SP(SP$_1$) $\to$ DM+2DAF (AFM$_1$) states,
which agrees with the variation in Fig.~\ref{fig1}. 
As for the case of chemical $P$, namely, 
 with the variation among different $X$ other than PF$_6$ and SbF$_6$, 
 our work suggests that a careful reconsideration for each compound should be made for
 how to allocate ``ambient $P$" positions, where the anisotropic parameters sensitively reflect the ground state; we leave them as a future problem.
Our results indicate that the dimensional crossover in magnetic states is controlled by not only $t_b$
but also $V$; inducing the FCO state
is essential to understand the sequence of phase transitions in TMTTF salts.
The apparently confusing cooperative behavior in the FCO and AFM$_2$ states
is naturally understood based on our scenario. 

We thank M. Ogata and Y. Otsuka for discussions and
K. Furukawa and T. Nakamura for providing us the crystal structure data.
This material is based upon work supported in part by
the National Science Foundation under Grants No. DMR-0804625 and No. DMR-1105531
and Grant-in-Aid for Scientific Research (No. 20110003, No. 20110004, and No. 21740270)
from MEXT, Japan.


\begin{thebibliography}{9}
\bibitem{Work_SF}
$|t_{q1}/t_b|$ is negligible for the PF$_6$ salt but not for the SbF$_6$ salt.
We confirm that it works as spin frustration between the interchain dimers in the DM state
but stabilizes the AFM state by releasing the frustration due to FCO
and therefore acts as enhancing the 2-$D$ in the spin sector seen in our results here.
\bibitem{comment_for_V}
Typical values for $V/U$ in this class of molecular conductors are estimated to be in a range of about $0.2-0.6$~\cite{Seo06}.
\bibitem{comment_size}
Previous studies show that the size dependence of the lattice modulations along the $a$-axis
is small beyond $8$ sites in the one-dimensional EHM~\cite{Clay03}.
\end{thebibliography}
\end{document}